\documentclass[aps,prd,superscriptaddress,preprint,tightenlines,nofootinbib]{revtex4}

\usepackage{graphicx}
\usepackage{dcolumn}
\usepackage{bm}

\newcommand{\text}{\mathrm}

\begin{document}

\preprint{CLNS 09/2055}
\preprint{CLEO 09-08}

\title{\boldmath $D^+_s$ Exclusive Hadronic Decays Involving $\omega$}

\author{J.~Y.~Ge}
\author{D.~H.~Miller}
\author{I.~P.~J.~Shipsey}
\author{B.~Xin}
\affiliation{Purdue University, West Lafayette, Indiana 47907, USA}
\author{G.~S.~Adams}
\author{D.~Hu}
\author{B.~Moziak}
\author{J.~Napolitano}
\affiliation{Rensselaer Polytechnic Institute, Troy, New York 12180, USA}
\author{K.~M.~Ecklund}
\affiliation{Rice University, Houston, Texas 77005, USA}
\author{Q.~He}
\author{J.~Insler}
\author{H.~Muramatsu}
\author{C.~S.~Park}
\author{E.~H.~Thorndike}
\author{F.~Yang}
\affiliation{University of Rochester, Rochester, New York 14627, USA}
\author{M.~Artuso}
\author{S.~Blusk}
\author{S.~Khalil}
\author{R.~Mountain}
\author{K.~Randrianarivony}
\author{T.~Skwarnicki}
\author{S.~Stone}
\author{J.~C.~Wang}
\author{L.~M.~Zhang}
\affiliation{Syracuse University, Syracuse, New York 13244, USA}
\author{G.~Bonvicini}
\author{D.~Cinabro}
\author{A.~Lincoln}
\author{M.~J.~Smith}
\author{P.~Zhou}
\author{J.~Zhu}
\affiliation{Wayne State University, Detroit, Michigan 48202, USA}
\author{P.~Naik}
\author{J.~Rademacker}
\affiliation{University of Bristol, Bristol BS8 1TL, UK}
\author{D.~M.~Asner}
\author{K.~W.~Edwards}
\author{J.~Reed}
\author{A.~N.~Robichaud}
\author{G.~Tatishvili}
\author{E.~J.~White}
\affiliation{Carleton University, Ottawa, Ontario, Canada K1S 5B6}
\author{R.~A.~Briere}
\author{H.~Vogel}
\affiliation{Carnegie Mellon University, Pittsburgh, Pennsylvania 15213, USA}
\author{P.~U.~E.~Onyisi}
\author{J.~L.~Rosner}
\affiliation{University of Chicago, Chicago, Illinois 60637, USA}
\author{J.~P.~Alexander}
\author{D.~G.~Cassel}
\author{R.~Ehrlich}
\author{L.~Fields}
\author{L.~Gibbons}
\author{S.~W.~Gray}
\author{D.~L.~Hartill}
\author{B.~K.~Heltsley}
\author{J.~M.~Hunt}
\author{J.~Kandaswamy}
\author{D.~L.~Kreinick}
\author{V.~E.~Kuznetsov}
\author{J.~Ledoux}
\author{H.~Mahlke-Kr\"uger}
\author{J.~R.~Patterson}
\author{D.~Peterson}
\author{D.~Riley}
\author{A.~Ryd}
\author{A.~J.~Sadoff}
\author{X.~Shi}
\author{S.~Stroiney}
\author{W.~M.~Sun}
\author{T.~Wilksen}
\affiliation{Cornell University, Ithaca, New York 14853, USA}
\author{J.~Yelton}
\affiliation{University of Florida, Gainesville, Florida 32611, USA}
\author{P.~Rubin}
\affiliation{George Mason University, Fairfax, Virginia 22030, USA}
\author{N.~Lowrey}
\author{S.~Mehrabyan}
\author{M.~Selen}
\author{J.~Wiss}
\affiliation{University of Illinois, Urbana-Champaign, Illinois 61801, USA}
\author{M.~Kornicer}
\author{R.~E.~Mitchell}
\author{M.~R.~Shepherd}
\author{C.~M.~Tarbert}
\affiliation{Indiana University, Bloomington, Indiana 47405, USA }
\author{D.~Besson}
\affiliation{University of Kansas, Lawrence, Kansas 66045, USA}
\author{T.~K.~Pedlar}
\author{J.~Xavier}
\affiliation{Luther College, Decorah, Iowa 52101, USA}
\author{D.~Cronin-Hennessy}
\author{K.~Y.~Gao}
\author{J.~Hietala}
\author{T.~Klein}
\author{R.~Poling}
\author{P.~Zweber}
\affiliation{University of Minnesota, Minneapolis, Minnesota 55455, USA}
\author{S.~Dobbs}
\author{Z.~Metreveli}
\author{K.~K.~Seth}
\author{B.~J.~Y.~Tan}
\author{A.~Tomaradze}
\affiliation{Northwestern University, Evanston, Illinois 60208, USA}
\author{S.~Brisbane}
\author{J.~Libby}
\author{L.~Martin}
\author{A.~Powell}
\author{C.~Thomas}
\author{G.~Wilkinson}
\affiliation{University of Oxford, Oxford OX1 3RH, UK}
\author{H.~Mendez}
\affiliation{University of Puerto Rico, Mayaguez, Puerto Rico 00681}
\collaboration{CLEO Collaboration}
\noaffiliation

\date{June 11, 2009}

\begin{abstract}
Using data collected near the $D^{\ast \pm}_s D^{\mp}_s$ peak production energy
$E_{\text{cm}} = 4170$~MeV by the CLEO-c detector, we search for
$D^+_s$ exclusive hadronic decays involving $\omega$. We find
$\mathcal{B}(D^+_s \rightarrow \pi^+ \omega) = (0.21 \pm 0.09 \pm 0.01) \%$,
$\mathcal{B}(D^+_s \rightarrow \pi^+ \pi^0 \omega) = (2.78 \pm 0.65 \pm 0.25) \%$,
$\mathcal{B}(D^+_s \rightarrow \pi^+ \pi^+ \pi^- \omega) = (1.58 \pm 0.45 \pm 0.09) \%$,
$\mathcal{B}(D^+_s \rightarrow \pi^+ \eta \omega) = (0.85 \pm 0.54 \pm 0.06)  \%$,
$\mathcal{B}(D^+_s \rightarrow K^+ \omega) <0.24 \%$,
$\mathcal{B}(D^+_s \rightarrow K^+ \pi^0 \omega) <0.82 \%$,
$\mathcal{B}(D^+_s \rightarrow K^+ \pi^+ \pi^- \omega) <0.54 \%$, and
$\mathcal{B}(D^+_s \rightarrow K^+ \eta \omega) <0.79 \%$. The upper 
limits are at 90\% confidence level.
\end{abstract}

\maketitle

The inclusive $\omega$ yield, $D^+_s \rightarrow \omega X$, is
substantial $(6.1\pm1.4)\%$~\cite{DsIn}. This is very surprising,
as the only $D^+_s$
exclusive decay mode involving $\omega$ that has been observed is
$D^+_s \rightarrow \pi^+ \omega $, with a branching fraction of
$\mathcal{B}(D^+_s \rightarrow \pi^+ \omega) =
(0.25\pm0.09)\%$~\cite{PDGValue}. There is lots of room for more $D^+_s$
exclusive hadronic decays involving $\omega$. The study of $\omega$
production in $D^+_s$ decays is of interest in shedding light on
mechanisms of weak decay and their interplay with long-distance
(nonperturbative) physics~\cite{Gronau:2009mp}.
Here we present a search for several $D^+_s$ exclusive hadronic decays
involving $\omega$. In particular, we consider final states with one,
two, and three pions: 
$D^+_s \rightarrow \pi^+ \omega$, 
$D^+_s \rightarrow \pi^+ \pi^0 \omega$, and
$D^+_s \rightarrow \pi^+ \pi^+ \pi^- \omega$. We also search for 
$D^+_s \rightarrow \pi^+ \eta \omega$, as this has been
suggested~\cite{Gronau:2009mp} as possibly being a large
decay mode. Finally, we search for modes in which one of the $\pi^+$
from the above-mentioned decays is replaced by a $K^+$:
$D^+_s \rightarrow K^+ \omega$, 
$D^+_s \rightarrow K^+ \pi^0 \omega$,
$D^+_s \rightarrow K^+ \pi^+ \pi^- \omega$, and
$D^+_s \rightarrow K^+ \eta \omega$. These last four modes would be
Cabibbo-suppressed, and hence are not expected to be large.

In this study we use 586 $\mathrm{pb}^{-1}$ of data produced in $e^+
e^-$ collisions at the Cornell Electron Storage Ring (CESR) and
collected by the CLEO-c detector near the center-of-mass (CM) energy
$\sqrt{s}=4170$~MeV. At this energy the cross-section for the channel
of interest, $D^{\ast +}_s D^-_s$ or $D^+_s D^{\ast -}_s$, is 
approximately 1 nb~\cite{Poling:2006da}. We select events in which the
$D^{\ast}_s$ decays to $D_s \gamma$ (94\% branching
fraction~\cite{PDGValue}). Other charm production totals $\sim$7
nb~\cite{Poling:2006da}, and the underlying light-quark ``continuum''
is about 12 nb.

The CLEO-c detector is a general-purpose solenoidal detector,
which is described in detail
elsewhere~\cite{Briere:2001rn,Kubota:1991ww,cleoiiidr,cleorich}. 
The charged particle tracking system covers a solid angle of 93\% of
$4 \pi$ and consists of a small-radius, six-layer, low-mass, stereo
wire drift chamber, concentric with, and surrounded by, a 47-layer
cylindrical central drift chamber. The chambers operate in a 1.0 T
magnetic field. The root-mean-square (rms) momentum resolution
achieved with the tracking system is approximately 0.6\% at
$p=1$~GeV/$c$ for tracks that traverse all layers of the drift
chamber.
Photons are detected in an electromagnetic calorimeter consisting of
about 7800 CsI(Tl) crystals~\cite{Kubota:1991ww}. The calorimeter
attains an rms photon energy resolution of 2.2\% at
$E_\gamma=1$~GeV and 5\% at 100~MeV. The solid angle coverage for
neutral particles in the CLEO-c detector is 93\% of $4 \pi$.
We utilize two particle identification (PID) devices to separate
charged kaons from pions: the central drift chamber, which provides
measurements of ionization energy loss ($dE/dx$), and, surrounding
this drift chamber, a cylindrical ring-imaging Cherenkov (RICH)
detector, whose active solid angle is 80\% of $4 \pi$. The combined
PID system has a pion or kaon efficiency $>85\%$ and a probability of
pions faking kaons (or vice versa) $<5\%$~\cite{CLEO:sys}. 
The detector response is modeled with a detailed
GEANT-based~\cite{geant} Monte Carlo (MC) simulation, with initial
particle trajectories generated by EvtGen~\cite{evtgen} and
final state radiation produced by PHOTOS~\cite{photos}.
The modeling of initial-state radiation is based on cross sections for
$D^{\ast \pm}_s D^\mp_s$ production at lower energies obtained from
the CLEO-c energy scan~\cite{Poling:2006da} near the CM energy where
we collected the sample. 

Here we employ a double-tagging technique, the same as the technique
that is used in the $D^+_s$ inclusive decay analysis~\cite{DsIn}.
Single-tag (ST) events are selected by fully reconstructing a
$D^{-}_{s}$, which we call a tag, in one of the following three
two-body hadronic decay modes: $D_s^- \to K^0_S K^-$, 
$D^-_s \to \phi \pi^-$ and $D^-_s \to K^{\ast 0} K^-$. (Mention of a
specific mode implies the use of the charge conjugate mode as well
throughout this Letter.) Details on the tagging selection procedure
are given in Refs.~\cite{DsIn, FanDspp, CLEO:taunu}.
The tagged $D^-_s$ candidate can be either the primary $D^-_s$ or the
secondary $D^-_s$ from the decay $D^{\ast -}_{s} \to \gamma D^-_s$.
We require the resonance decays to satisfy the following mass
windows around the nominal masses~\cite{PDGValue}: $K^0_S \to \pi^+
\pi^-$ ($\pm 12$~MeV), $\phi \to K^+ K^-$ ($\pm 10$~MeV) and
$K^{\ast 0} \to K^+ \pi^-$ ($\pm 75$~MeV). The momenta of all charged
particles utilized in tags are required to be 100~MeV/$c$ or greater
to suppress the slow pion background from $D^\ast \bar{D}^\ast$ and
$D^\ast D$ decays
(through $D^\ast \to \pi D$).

The reconstructed invariant mass of the $D_s$ candidate,
$M(D_s)$, and the mass recoiling against the $D_s$ candidate, 
$M_\text{recoil}(D_s)
\equiv \sqrt{ (E_{0} - E_{D_s} )^2 - (\mathbf{p}_{0}-\mathbf{p}_{D_s})^2 }
$, are used as our primary kinematic variables to select a $D_s$
candidate. Here $(E_{0},\mathbf{p}_{0})$ is the net four-momentum of
the $e^+e^-$ beams, taking the finite beam crossing angle into account,
$\mathbf{p}_{D_s}$ is the momentum of the $D_s$ candidate, 
$E_{D_s} = \sqrt{m^2_{D_s} + \mathbf{p}^2_{D_s}}$,
and $m_{D_s}$ is the known $D_s$ mass~\cite{PDGValue}.
We require the recoil mass to be within $55$ MeV of the $D^\ast_s$
mass~\cite{PDGValue}. This loose window allows both primary and
secondary $D_s$ tags to be selected. We also require a photon
consistent with coming from $D^{\ast}_s \rightarrow \gamma D_s$
decay, by looking at the mass recoiling against the $D_s$ candidate
plus $\gamma$ system, $M_\text{recoil}(D_s \gamma)
\equiv \sqrt{ (E_{0} - E_{D_s} - E_\gamma)^2 - (\mathbf{p}_{0}-\mathbf{p}_{D_s}-\mathbf{p}_\gamma)^2 }$.
For correct combinations, this recoil mass peaks at $m_{D_s}$,
regardless of whether the candidate is due to a primary or a secondary
$D_s$. We require 
$| M_\text{recoil}(D_s \gamma) - m_{D_{s}} | < 30~\text{MeV}$.

\begin{figure*}
  \centering{
    \includegraphics*[width=0.95\textwidth]{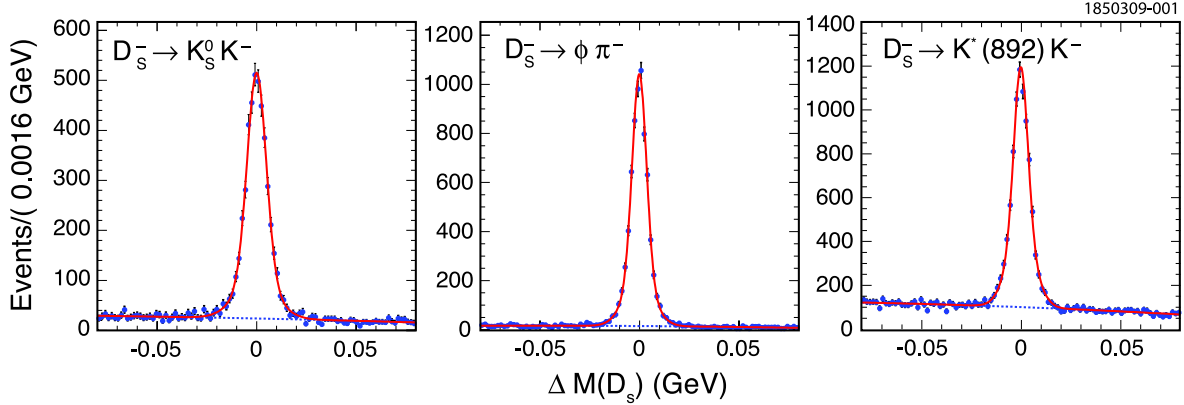} 
  }
  \caption{The mass difference $\Delta M (D_s) \equiv M(D_s) - m_{D_s}$
    distributions in each tag mode.
    We fit the $\Delta M(D_s)$ distribution (points)
    to the sum (solid curve)
    of signal (double-Gaussian)
    plus background (second degree polynomial, dashed curve)
    functions.}
  \label{fig:tag}
\end{figure*}

The distributions of $\Delta M(D_s) \equiv M(D_s) - m_{D_s}$, the
invariant mass difference of $D_s$ tag candidates, after
applying the two recoil mass requirements for each tag mode are shown
Fig.~\ref{fig:tag}. We use the tag invariant mass sidebands to
estimate the backgrounds in our signal yields from the wrong tag
combinations (incorrect combinations that, by chance, lie within the
$\Delta M(D_s)$ signal region). The signal region is
$|\Delta M(D_s)| < 20$~MeV, while the sideband region is 
$35$~MeV $< |\Delta M(D_s)| < 55$~MeV. To find the sideband scaling
factor, the $\Delta M(D_s)$ distributions are fit to the sum of
double-Gaussian signal plus second-degree polynomial background
functions. We have 18586~$\pm$~163 ST events that we use for further
analysis. 

In each event where a tag is identified, we search for our signal
modes recoiling against the tag. Charged tracks utilized in
signal candidates are required to satisfy criteria based on the track
fit quality, have momenta above 50~MeV/$c$, and angles $\theta$ with
respect to the beam line, satisfying $|\cos\theta|<0.93$. They must
also be consistent with coming from the interaction point in three
dimensions. Pion and kaon candidates are required to have $dE/dx$
measurements within three standard deviations ($3\sigma$) of the
expected value. For tracks with momenta greater than 700~MeV/$c$, RICH
information, if available, is combined with $dE/dx$.

We identify $\pi^{0}$
candidates via $\pi^{0} \rightarrow \gamma \gamma$, detecting the
photons in the CsI calorimeter. To avoid having both photons in a
region of poorer energy resolution, we require that at least one of
the photons be in the ``good barrel'' region, $|\cos \theta_{\gamma}|
< 0.8$. We require that the calorimeter clusters have a measured
energy above 30~MeV, have a lateral distribution consistent with that
from photons, and not be matched to any charged track. The invariant
mass of the photon pair is required to be within 3$\sigma$
($\sigma\sim$ 6 MeV) of the known $\pi^0$ mass. The $\eta$ candidates
are formed using a similar procedure as for $\pi^0$ except that
$\sigma\sim$ 12~MeV. The $\pi^0$ and $\eta$ mass constraints are
imposed when $\pi^0$ or $\eta$ candidates are used in further
reconstruction. We reconstruct $\omega$ candidates in the $\omega
\rightarrow \pi^+ \pi^- \pi^0$ decay mode. 

Mode-dependent requirements on numbers of charged kaons and pions are
applied on the signal side. For example, we require there must be
exactly one charged kaon and two charged pions for the $D^+_s
\rightarrow K^+ \omega, \omega \rightarrow \pi^+ \pi^- \pi^0$ mode. No
extra tracks are allowed on the signal side. The best $\pi^0$ (or
$\eta$) candidate is selected based on the pull mass (number of
standard deviations of measured mass from true mass). For $D^{+}_{s}
\rightarrow \pi^+\eta\omega$, a veto on $\eta' \rightarrow \pi^+ \pi^-
\eta$ has been applied to remove the dominant background contribution
from the $D^{+}_{s} \rightarrow \pi^+\pi^0 \eta'$ decay.

The double-tag (DT) yields are extracted from the $\pi^+ \pi^- \pi^0$
invariant mass distribution after requiring that both the tagging
$D_s$ and signal $D_s$ invariant masses be in the $D_s$ nominal mass
region ($20$~MeV mass window on the tag side and $30$~MeV mass window
on the signal side due to $\pi^0$ or $\eta$ on the signal side). The
$\omega$ mass signal region is $|M_{\pi^+\pi^-\pi^0}-m_{\omega}| <
20$~MeV, while the sideband region is $40$~MeV $<
|M_{\pi^+\pi^-\pi^0}-m_{\omega}| < 80$~MeV, where
$M_{\pi^+\pi^-\pi^0}$ is the $\pi^+\pi^-\pi^0$ invariant mass and
$m_{\omega}$ is the nominal mass of $\omega$~\cite{PDGValue}. 

\begin{figure*}
  \centering{
    \includegraphics*[width=0.95\textwidth]{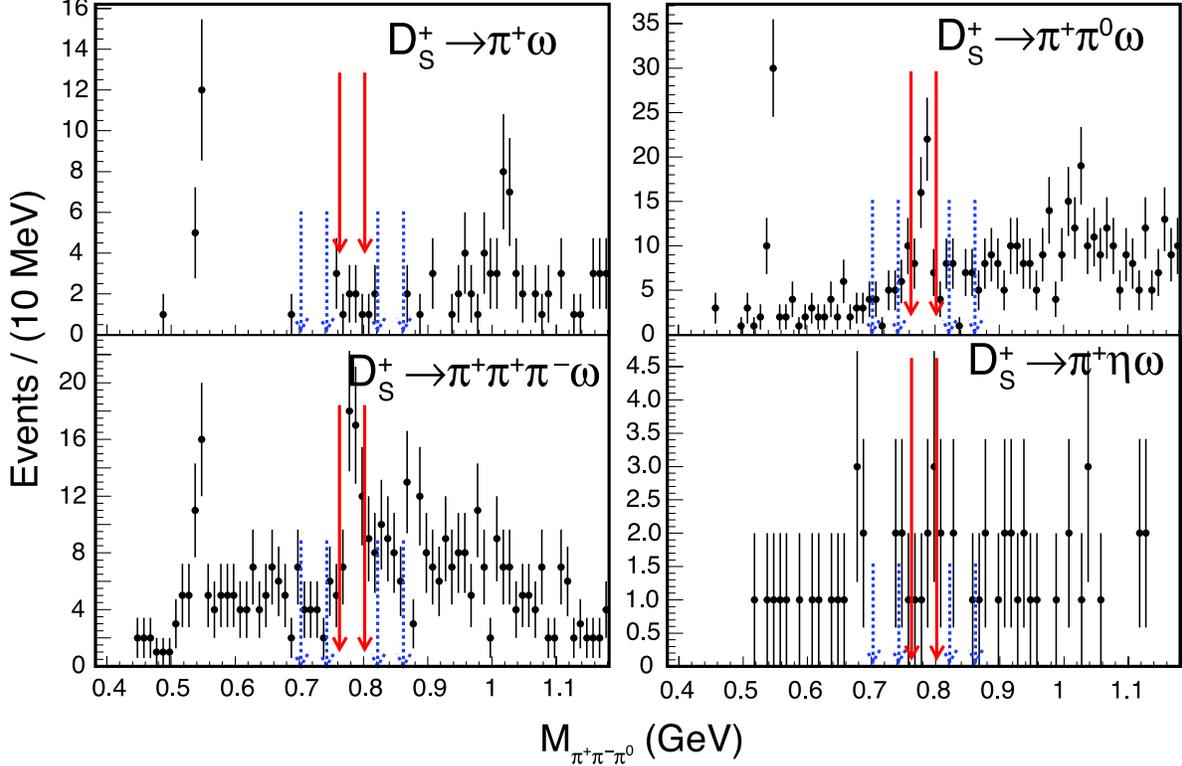}}
  \caption{Invariant mass distributions of $\omega$ candidates for the
    first four modes: $D^{+}_{s} \rightarrow \pi^+\omega$,
    $D^{+}_{s} \rightarrow \pi^+\pi^0\omega$,  $D^{+}_{s} \rightarrow
    \pi^+\pi^+\pi^-\omega$, and  $D^{+}_{s} \rightarrow \pi^+\eta\omega$.
    The solid lines (red online) indicate the $\omega$ mass signal region and
    the dashed lines (blue online) indicate the $\omega$ mass sideband
    regions. Peaks from $\eta \rightarrow \pi^+ \pi^- \pi^0$ and $\phi
    \rightarrow \pi^+ \pi^- \pi^0$ are also evident.} 
  \label{fig:omgmassa}
\end{figure*}

The invariant mass distributions of $\omega$ candidates for the first
four modes are shown in Fig.~\ref{fig:omgmassa} and the
yields are given in Table~\ref{table:yieldsr}. For $D^{+}_{s}
\rightarrow \pi^+\omega$, $D^{+}_{s} \rightarrow \pi^+\pi^0\omega$,
and $D^{+}_{s} \rightarrow \pi^+ \pi^+\pi^-\omega$, clear signals are
found in the data. Note also the peaks from $\eta \rightarrow \pi^+
\pi^- \pi^0$ and $\phi \rightarrow \pi^+ \pi^- \pi^0$, corresponding
to known $D_s$ decays to $\eta$ plus pions and $\phi$ plus pions.

For $D^{+}_{s} \rightarrow
\pi^+\eta\omega$, we observe 7 events in the signal region, and 5 events
in a sideband region twice as wide. From Binomial statistics, the
probability of observing 7 or more events in the signal region and 5
or fewer events in the sideband region, out of a total of 12 events,
if there is no true signal, is
$6.6\%$. Thus we have ``evidence'' for $D^{+}_{s} \rightarrow
\pi^+\eta\omega$, but cannot claim ``observation''. We quote both a
value for the branching fraction and an upper limit on it.

We find no significant evidence for any of the modes with kaons, and
therefore set upper limits on their branching fractions. The numbers
of events from $\omega$ mass signal and sideband regions for these
four modes are given in Table~\ref{table:yieldsr}. Upper limits are
calculated using Poisson statistics, and allowing for uncertainty in
number of background events.

\begin{table}
  \centering
  \caption{\label{table:yieldsr}Observed yields. Here
    $N_{\text{Sg}}$ is the observed event number from $\omega$ mass
    signal region, $N_{\text{Sd}}$ is the scaled event number from
    $\omega$ mass sideband regions, and $N_{\text{Ss}}$ is the
    sideband-subtracted signal yield. Double-tag detection
    efficiencies ($\epsilon_{\text{DT}}$) are listed in the last
    column. The efficiencies include sub-mode branching
    fractions~\cite{PDGValue}, and have been corrected to include
    several known small differences between data and Monte Carlo
    simulation. Errors shown are statistical errors only.}
  \begin{tabular}{l c c c c } \hline \hline
    Mode 
    & ~~$N_{\text{Sg}}$~~ 
    & ~~$N_{\text{Sd}}$~~ 
    & ~~$N_{\text{Ss}}$~~
    & ~~~~$\epsilon_{\text{DT}}$(\%)~~~~\\ \hline
    $D^+_s \rightarrow \pi^+ \omega$
    &6.0 &0.0 &6.0 $\pm$2.4&4.07$\pm$0.08\\
    $D^+_s \rightarrow \pi^+ \pi^0 \omega$      
    &53.0&19.0&34.0$\pm$7.9&1.75$\pm$0.04\\ 
    $D^+_s \rightarrow \pi^+ \pi^+ \pi^- \omega$
    &54.0&24.8&29.2$\pm$8.2&2.64$\pm$0.07\\
    $D^+_s \rightarrow \pi^+ \eta \omega$       
    &7.0 &2.5 &4.5 $\pm$2.9&0.76$\pm$0.04\\
    $D^+_s \rightarrow K^+ \omega$                 
    &3.0 &2.0 &1.0 $\pm$2.0&3.66$\pm$0.08\\
    $D^+_s \rightarrow K^+ \pi^0 \omega$           
    &4.0 &2.5 &1.5 $\pm$2.3&1.32$\pm$0.05\\
    $D^+_s \rightarrow K^+ \pi^+ \pi^- \omega$     
    &3.0 &1.5 &1.5 $\pm$1.9&1.72$\pm$0.05\\
    $D^+_s \rightarrow K^+ \eta\omega$             
    &0.0 &0.0 &0.0 $\pm$0.0&0.45$\pm$0.03\\
    \hline \hline
  \end{tabular}
\end{table}

We determine the efficiency for detecting our tags
($\epsilon_{\text{ST}}$) using Monte Carlo samples in which one $D_s$
decays into the tag mode, and the other decays generically. The
single-tag efficiency weighted over the three tags is (26.63 $\pm$
0.14)\%. We determine the efficiency for detecting both tag and signal
($\epsilon_{\text{DT}}$) using Monte Carlo samples in which one $D_s$
decays into a tag mode, and the other $D_s$ decays into a signal
mode. For the three-body decay modes, we assume three-body phase 
space. For $D^+_s \rightarrow \pi^+ \pi^0 \omega$, we also consider
$D^+_s \rightarrow \rho^+ \omega, \rho^+ \rightarrow \pi^+ \pi^0$. For
$D^+_s \rightarrow K^+ \pi^0 \omega$, we also consider $D^+_s
\rightarrow K^{\ast +} \omega, K^{\ast +} \rightarrow K^+ \pi^0$. For
$D^+_s \rightarrow \pi^+ \pi^+ \pi^- \omega$, we assume $\pi^+ \pi^+
\pi^- \omega$ phase space and estimate a systematic error by using
$D^+_s \rightarrow \pi^+ \rho^0 \omega, \rho^0 \rightarrow \pi^+
\pi^-$. Double-tag detection efficiencies are given in
Table~\ref{table:yieldsr}.

The absolute branching fractions ($\mathcal{B}_\text{mode}$) are
obtained from the ST yield ($N_\text{ST}$) and DT yield
($N_\text{DT}$) without needing to know the integrated luminosity or
the produced number of $D^{\ast \pm}_s D^\mp_s$ pairs,
\begin{equation}
  \mathcal{B}_\text{mode} = \frac{N_\text{DT}}
  {N_\text{ST}} \times
  \frac{\epsilon_\text{ST}}{\epsilon_\text{DT}}.
\end{equation}

\begin{figure}
  \centering{
    \includegraphics*[width=0.45\textwidth]{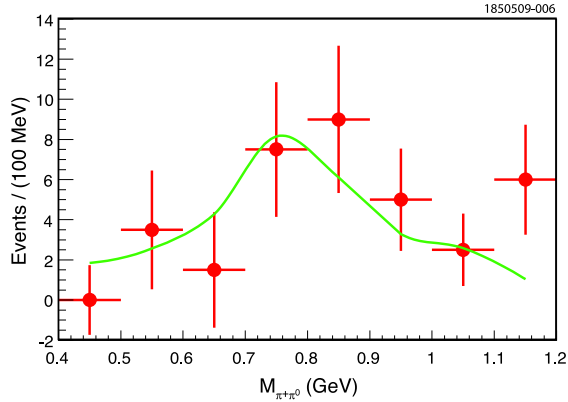}}
  \caption{$\pi^+\pi^0$ invariant mass distribution. We fit to the sum
    of phase space $\pi^+\pi^0\omega$ MC and $\rho^+\omega$ MC. The
    fit result suggests that $0.52\pm0.30$ of the $D^+_s \rightarrow
    \pi^+ \pi^0 \omega$ decay come from the $D^+_s \rightarrow \rho^+
    \omega, \rho^+ \rightarrow \pi^+ \pi^0$ decay. The points are the
    data and the superimposed line is the fit.}
  \label{fig:fitrho}
\end{figure}

The $D^+_s \rightarrow \pi^+ \pi^0 \omega$ decay might come from 
$D^+_s \rightarrow \rho^+(\pi^+ \pi^0) \omega$. We fit the
corresponding $\pi^+\pi^0$ invariant mass distribution to the sum of
phase space $\pi^+\pi^0\omega$ MC and $\rho^+\omega$ MC. The fit 
result suggests that $0.52\pm0.30$ of the $D^+_s \rightarrow \pi^+ \pi^0
\omega$ decay come from the $D^+_s \rightarrow \rho^+ \omega, \rho^+
\rightarrow \pi^+ \pi^0$ decay, as shown in Fig.~\ref{fig:fitrho}.

We have considered several sources of systematic uncertainty. The
uncertainty associated with the efficiency for finding a track is
0.3\%; an additional 0.6\% systematic uncertainty for each kaon track
is added~\cite{CLEO:sys}. The relative systematic uncertainties for
$\pi^0$ and $\eta$ efficiencies are 4.0\%. Uncertainties in the
charged pion and kaon identification efficiencies are 0.3\% per pion
and 0.3\% per kaon~\cite{CLEO:sys}. For the $D^+_s \rightarrow
\pi^+\pi^0\omega$ mode, the relative contribution of $D^+_s
\rightarrow \rho^+\omega, \rho^+ \rightarrow \pi^+\pi^0$ is determined
to be $0.52 \pm0.30$ from the fit. We use the central value of that
ratio to calculate the efficiency and take the error as a systematic
uncertainty. All Monte Carlo efficiencies have been corrected to
include several known small differences between data and Monte Carlo
simulation. Upper limits have been increased to allow for the
systematic errors.

\begin{table}
  \centering
  \caption{\label{table:result}Branching fractions and upper
    limits. Uncertainties are statistical and systematic,
    respectively.}
  \begin{tabular}{l  c }
    \hline \hline
     Mode  &  $\mathcal{B}_\text{mode}(\%)$ \\ \hline\hline
    $D^+_s \rightarrow \pi^+ \omega$~~~~~~~~~~~~~~~~~~~~~~~~
    &$0.21 \pm 0.09 \pm 0.01$\\
    $D^+_s \rightarrow \pi^+ \pi^0 \omega$
    &$2.78 \pm 0.65 \pm 0.25$\\
    $D^+_s \rightarrow \pi^+ \pi^+ \pi^- \omega$
    &$1.58 \pm 0.45 \pm 0.09$\\
    $D^+_s \rightarrow \pi^+ \eta \omega$
    &$0.85 \pm 0.54 \pm 0.06$\\ 
    &$<2.13~(90\%~\text{CL})$\\
    $D^+_s \rightarrow K^+ \omega$
    &$<0.24~(90\%~\text{CL})$\\
    $D^+_s \rightarrow K^+ \pi^0 \omega$
    &$<0.82~(90\%~\text{CL})$\\
    $D^+_s \rightarrow K^+ \pi^+ \pi^- \omega$
    &$<0.54~(90\%~\text{CL})$\\
    $D^+_s \rightarrow K^+ \eta\omega$
    &$<0.79~(90\%~\text{CL})$\\
   \hline\hline
  \end{tabular}
\end{table}

The branching fractions and upper limits are listed in
Table~\ref{table:result}. In summary, we report first observations of
$D^+_s \rightarrow \pi^+\pi^0\omega$ and $D^+_s \rightarrow
\pi^+\pi^+\pi^-\omega$ decays. The branching fractions are
substantial. We find evidence for the $D^+_s \rightarrow
\pi^+\eta\omega$ decay. Our measurement of $D^+_s \rightarrow \pi^+
\omega$ decay is in good agreement with the PDG value~\cite{PDGValue},
and of comparable accuracy. The sum of branching fractions of these
four observed modes is (5.4$\pm$1.0)\%, which accounts for most of the
$D_s$ inclusive $\omega$ decays (6.1$\pm$1.4)\%~\cite{DsIn}. We also
report the first upper limits on $D^+_s \rightarrow K^+\omega$, $D^+_s
\rightarrow K^+\pi^0\omega$, $D^+_s \rightarrow K^+\pi^+\pi^-\omega$,
and $D^+_s \rightarrow K^+\eta\omega$ decays.

We gratefully acknowledge the effort of the CESR staff
in providing us with excellent luminosity and running conditions.
This work was supported by
the A.P.~Sloan Foundation,
the National Science Foundation,
the U.S. Department of Energy,
the Natural Sciences and Engineering Research Council of Canada, and
the U.K. Science and Technology Facilities Council.

\end{document}